\documentclass[a4paper,10pt, twocolumn]{article}
\usepackage[utf8]{inputenc}
\usepackage[francais]{babel}
\usepackage{graphicx,version,amsmath}
\usepackage{pstricks,pst-node,pst-plot}
\usepackage[cdot,squaren,Gray,textstyle]{SIunits}
\usepackage{parskip}
\usepackage{titlesec}

\def\arxivdegrees{\leavevmode\hbox to 0.3em{\hss\degre\hss}}

\titleformat{\section}%
	{\normalfont\Large\bfseries }{\thesection}{1cm}{}
\titleformat{\subsection}%
	{\normalfont\large\bfseries }{\thesection}{1cm}{}
\titlespacing{\section}{0pt}{*1}{*-1}
\titlespacing{\subsection}{0pt}{*1}{*0}

\begin{document}

\date{}% no date on the front page!

\title{\vspace{-8mm}\textbf{\large
	Synthèse sonore de clarinette avec modèle %
	de résonateur à trous latéraux.}}

\author{
Fabrice Silva$^1$, Philippe Guillemain$^1$, %
 Jean Kergomard$^1$ et Jean-Pierre Dalmont$^2$\\
$^1$ \emph{\small Laboratoire de Mécanique et d'Acoustique UPR CNRS 7051, %
31 Ch. J. Aiguier, 13402 Marseille cedex 20, France, 
% Marseille, France,
}\\
$^2$ \emph{\small Laboratoire d'Acoustique de l'Université du Maine UMR CNRS 6613, %
Av. O. Messiaen, 72085 Le Mans cedex 9, France.}\\
\emph{\small courriel : silva@lma.cnrs-mrs.fr}
}

\maketitle
\section*{Résumé}
\label{sec:Resume}
Les méthodes de synthèse sonore par modèles physiques reposent sur une description des divers éléments constituant l'instrument. Les contraintes liées à la synthèse temps-réel requièrent l'emploi de géométries simples pour le résonateur. Cependant le corps de la clarinette est bien plus complexe qu'un simple tuyau parfaitement cylindrique du fait notamment de la présence des trous latéraux. Nous présentons ici une manière de prendre en compte de façon globale et simplifiée l'ouverture d'un certain nombre de trous latéraux dans le cadre de la synthèse sonore en temps-réel.

\section*{Introduction}
\label{sec:Introduction}
Les méthodes de synthèse sonore reposant sur les modèles physiques nécessitent, dans un premier temps, de s'attacher à observer et comprendre quels sont les phénomènes les plus importants dans le mécanisme de production du son. Il ne s'agit pas de chercher à donner une description exhaustive des particularités de l'ensemble de l'instrument, mais de s'assurer que les caractéristiques des sons synthétisés soient proches de celles des sons de l'instrument réel. Pour la clarinette, un premier modèle simplifié de résonateur est un simple cylindre dont la longueur correspond à la distance entre le bec et le premier trou ouvert. 
%En ouvrant (et en fermant) successivement les trous latéraux, on modifie la longueur efficace de la colonne d'air et les fréquences de ses modes. De nombreux éléments peuvent agir comme des corrections de longueur, au moins en basses fréquences. Toutefois, certaines irrégularités peuvent engendrer des effets plus complexes.
 Partant de l'observation d'une différence significative  entre les mesures d'impédances et le modèle cylindrique, %de brillance des sons réel et de synthèse
nous nous sommes intéressés à la manière dont la prise en compte de la partie du corps de l'instrument placée en aval du premier trou ouvert modifie les propriétés acoustiques du résonateur.
\par Contrairement à d'autres méthodes où chacun des trous latéraux est pris en compte de manière détaillée (van Walstijn \cite{vWC03} et Scavone \cite{SC98}), nous nous sommes attachés à modéliser globalement l'effet de l'ensemble des trous ouverts en nous appuyant sur les travaux de Benade \cite{Ben76} et Kergomard \cite{Ker81,Ker89}. Une formulation temporelle numérique du comportement du réseau de trous latéraux ouverts a été élaborée afin de raffiner les modèles de résonateurs déjà mis en \oe uvre dans des algorithmes de synthèse sonore en temps-réel par modèle physique (Guillemain \cite{GKV05}). 
%\textbf{Parler de la différence de timbre entre le son de la clarinette réelle et le son synthétisé avec un cylindre simple, mentionner comme cause éventuelle possible la présence d'un guide en aval du trou ouvert. Citer les travaux de Scavone \cite{SC98} et vanWalstijn \cite{} pour une synthèse complète avec description précise de chaque élément. Expliquer le pourquoi d'une modélisation plus globale, partant des travaux de Benade \cite{Ben76} (modélisation par un réseau régulier), Kergomard \cite{Ker81} et Keefe \cite{Kee90}.}

\section*{Modèle physique du résonateur}
\begin{figure}[hbt]
\begin{center}
\includegraphics[width=9cm]{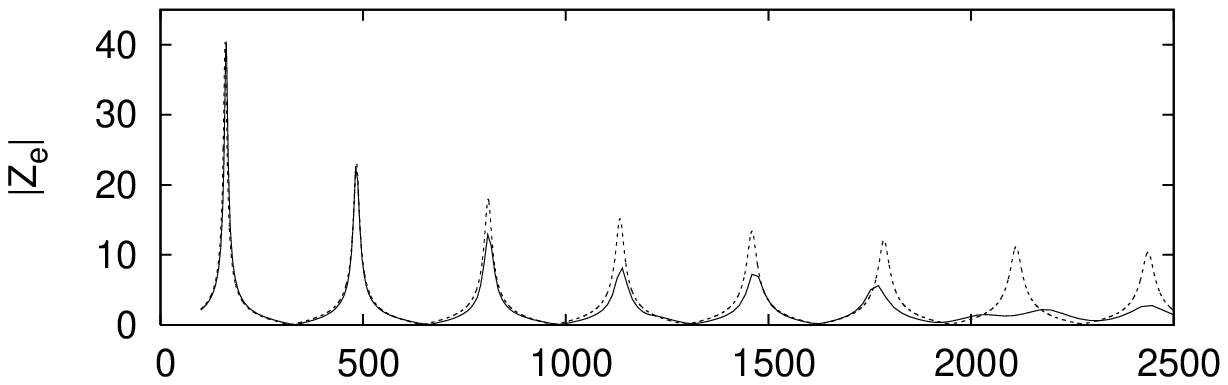}
\includegraphics[width=9cm]{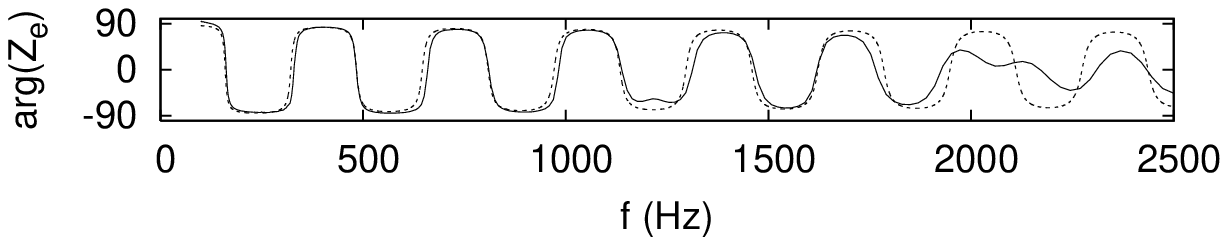}
\end{center}
\caption{Impédances d'entrée réduites, mesurée sur une clarinette réelle pour un doigté de $Fa\#_2$ (en trait plein) et calculée à partir d'un modèle de résonateur cylindrique (en pointillés).}
\label{fig:mesurecalage}
\end{figure}

%\subsection*{Considérations générales}
%\label{ssec:Hypotheses}
L'étude qui suit se place dans le domaine de l'acoustique linéaire : les mesures d'impédance sur lesquelles sont visibles les différences de comportement entre le corps d'une clarinette et un simple tube cylindrique ont été présentées par Benade \cite{Ben76}. Elles ont été mesurées à faible niveau : les différences ne proviennent pas d'éventuelles non-linéarités. La figure~\ref{fig:mesurecalage} montre un exemple typique de mesures réalisées dans le cadre de cette étude et une impédance calculée à partir d'un modèle cylindrique. Au vu des dimensions et du contenu spectral du son de l'instrument, nous ne considérons que la propagation d'ondes planes dans la colonne d'air : la fréquence d'apparition du premier mode non plan pour un conduit cylindrique de rayon $r=7\milli\metre$ se trouve en effet aux alentours de $14\kilo\hertz$ ce qui est bien en deçà des fréquences de résonance principales du résonateur. Sous ces hypothèses, la théorie de Kirchhoff permet de prendre en compte la viscosité et les effets thermiques dans le nombre d'onde $k$. Le rayon du guide d'onde étant bien plus grand que les épaisseurs des couches limites de pertes visco-thermiques $l_v$ et $l_t$, la constante de propagation est approchée classiquement par :
%\cite{Pie81} 
\begin{equation}
\label{eq:nbonde}
k=\frac\omega c%
	-j^{3/2}\frac{\sqrt{l_v}+\left(C_{pv}-1\right)\sqrt{l_t}}{r}%
	\sqrt{\frac{\omega}{c}}
\end{equation}
avec $c=340\metre\per\second$, $l_v=4\times\power{10}{-8}\metre$, $l_t=5.6\times\power{10}{-8}\metre$ et $C_{pv}=1.4$ (rapport des chaleurs spécifiques à pression et volume constant).
\vspace{-1cm}
\subsection*{Lignes de transmission}
\label{ssec:lignes}
\par Ces différentes hypothèses nous permettent d'adopter le formalisme de ligne de transmission dans le domaine fréquentiel (avec une dépendance temporelle en $\exp{j\omega t}$). Pour un cylindre, pression et débit acoustiques en entrée $x=0$ s'expriment en fonction des valeurs en sortie $x=L$ pour une pulsation $\omega$ donnée :
\begin{multline}
  \begin{pmatrix} P_e(\omega)\\Z_c U_e(\omega) \end{pmatrix}
  =M(L)\begin{pmatrix}P_s(\omega)\\Z_c U_s(\omega) \end{pmatrix}
  \\\mbox{ où }
  M(L)=\begin{pmatrix}\cos{(kL)}&j\sin{(kL)}\\
    j\sin{(kL)}&\cos{(kL)}\end{pmatrix} 
\end{multline}
où $Z_c=\rho c\,(\pi r^2)^{-1}$ est l'impédance caractéristique de la colonne d'air cylindrique, avec $\rho$ masse volumique de l'air. La dépendance de l'impédance d'entrée $Z_e$ du tuyau à sa charge $Z_s$ s'en déduit aisément :
\begin{equation}
\label{eq:zecylindrecharge}
Z_e(\omega)=\frac{P_e(\omega)}{U_e(\omega)}=\frac{Z_s(\omega)+jZ_c\tan{(kL)}}%
	{1+jZ_s(\omega)Z_c^{-1}\tan{(kL)}}.
\end{equation}

\subsection*{De la cellule élémentaire\ldots}
\label{ssec:cellule}
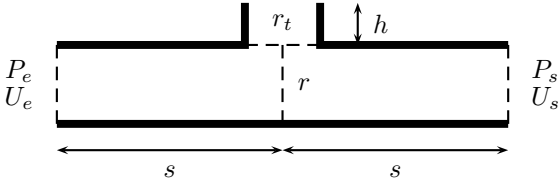
\begin{figure}[hbt]
    \begin{center}
    \psset{yunit=.7}
    \begin{pspicture}(0,-.5)(4,3)
	%\psaxes[labels=none,ticks=x,dx=2]{-}(0,0)(-.5,0)(4.5,0)
	\pcline{<->}(-1,0)(2,0)\Bput{$s$}
	\pcline{<->}(2,0)(5,0)\Bput{$s$}
	\pcline{<->}(3,2)(3,2.8)\Bput{$h$}
	\psline[linewidth=3pt]{-}(-1,0.5)(5,0.5)
	\psline[linewidth=3pt]{-}(-1,2)(1.5,2)(1.5,2.8)
	\psline[linewidth=3pt]{-}(2.5,2.8)(2.5,2)(5,2)
	\pcline[linestyle=dashed]{-}(1.5,2)(2.5,2)\Aput{$r_t$}
	\psline[linestyle=dashed]{-}(-1,.5)(-1,2)
	\pcline[linestyle=dashed]{-}(2,.5)(2,2)\Bput{$r$}
	\psline[linestyle=dashed]{-}(5,.5)(5,2)
	\rput(-1.5,1.5){$P_e$}\rput(-1.5,1){$U_e$}
	\rput(5.5,1.5){$P_s$}\rput(5.5,1){$U_s$}
	\end{pspicture}
    \end{center}
    \caption{Schéma d'une cellule élémentaire : notations.}
    \label{fig:cellule}
\end{figure}

Afin de parvenir à une modélisation globale d'un réseau de trous latéraux ouverts, nous nous intéressons en premier lieu au comportement d'une cellule élémentaire (Figure~\ref{fig:cellule}) comportant un tronçon de longueur $2s$ de conduit principal (de rayon $r$) muni d'une cheminée latérale de rayon $r_t$ et de hauteur $h_c$ incluant les corrections de longueurs aux deux extrémités de la longueur $h$ \cite{DKK+99}. 
%% Figure : schéma cellule

La matrice de transfert entre les grandeurs en entrée et en sortie de la cellule correspond donc aux propagations sur les éléments de longueur $s$ en amont et en aval de la cheminée, ainsi qu'à la dérivation au niveau du trou (conservation du débit et de la pression) :
\begin{equation}
\begin{pmatrix} P_e\\Z_c U_e\end{pmatrix}
	=M(s)\begin{pmatrix} 1 & 0\\Y_t & 1\end{pmatrix}M(s)
	\begin{pmatrix}P_s\\Z_c U_s\end{pmatrix},
\end{equation}
où $Y_t^{-1}=jZ_c kh_c\left({r}/{r_t}\right)^2$ est l'admittance présentée par la cheminée à la colonne d'air principale en supposant que les dimensions du trou sont petites devant la longueur d'onde (nous ignorons l'impédance en série liée au champ antisymétrique dans la cheminée).
\begin{comment}
En comparant le comportement d'un résonateur comprenant un cylindre de longueur $L$  suivi d'une cellule (soit une longueur de l'entrée jusqu'au trou de $L+s$) à celui d'un tube présentant une première fréquence de résonance identique, on se rend compte que la présence d'un guide d'onde en aval du trou ouvert modifie l'allure de la courbe d'impédance d'entrée (Figure~\ref{fig:guideaval}).
\end{comment}
%% Figure : effet du conduit aval
%% Courbe cylindre jusqu'au trou
%% 		 cylindre+1cellule

\subsection*{\ldots au réseau de trous latéraux ouverts}
\label{ssec:reseau}
La configuration étudiée ci-dessus ne s'applique qu'à un nombre limité de doigtés -- ceux pour lesquels un seul trou est ouvert. Pour les autres, Benade \cite{Ben76} a montré expérimentalement qu'il est possible de considérer le réseau périodique, les trous étant identiques et équidistants, ce que nous discuterons plus loin. Il existe alors pour le réseau de trous ouverts, selon les fréquences, des bandes d'arrêt et des bandes passantes. Dans les bandes d'arrêt, les ondes du réseau sont évanescentes, et l'impédance d'entrée du réseau est en pratique égale à son impédance itérative dès lors qu'il y a plus de deux cellules. Ceci vaut en particulier pour les basses fréquences. Au-dessus d'une fréquence de coupure, qui correspond, pour des raisons de symétrie, à la fréquence propre de la cellule fermée à ses deux extrémités, on a une bande passante. Dans celle-ci, le rayonnement devient très efficace, et les ondes sont fortement atténuées, ce qui entraîne aussi pour l'impédance d'entrée du réseau d'être proche de l'impédance itérative. Benade \cite{Ben76} avait bien noté ce point important, Kergomard \cite{Ker89} l'a interprété en l'associant à l'interaction extérieure des trous latéraux : une impédance d'entré de clarinette mesurée telle que représentée figure~\ref{fig:mesurecalage} montre de fait qu'il n'y a plus de résonances au-dessus de la coupure (environ $2\kilo\hertz$). \par Sous cette condition de forte atténuation et si l'on admet que le réseau est périodique, l'impédance d'entrée est l'impédance itérative \cite{Ker81} :
\begin{comment}
 on imagine qu'il suffit de disposer en cascade le nombre de cellule adapté. L'impédance d'entrée d'une association de $n$ cellules élémentaires identiques s'exprime aisément à l'aide de la constante de propagation $\phi$ et de l'impédance itérative~$Z_l$ :
\begin{gather}
\label{eq:ncellules}
Z_n=jZ_l\tan{(n\phi)}\,,\\
\cos{\phi} = \cos{2ks}+j\frac{Y_tZ_c}{2}\sin{2ks}\,,\\
Z_l = Z_c\sqrt{%
	\frac{\displaystyle 2+jY_tZ_c\tan{ks}}%
	{\displaystyle 2-jY_tZ_c\cot{ks}}}\,.
\end{gather}
si la dernière cellule voit une charge nulle.
\par \`A partir d'une nombre relativement faible de cellules, l'association finie se comporte comme si le réseau est infini. En effet, en présence d'atténuation (pertes dans les couches limites et rayonnement par les ouvertures) \cite{Ker89,Ker81}, $Z_n$ converge relativement vite vers $Z_l$. Ce phénomène est liée à la nature même de l'impédance itérative : si elle est placée à la sortie d'un élément, alors l'impédance d'entrée de la cellule chargée est égale à l'impédance itérative. Cette convergence est d'autant plus rapide que l'ultime cellule  débouche sur un pavillon exponentiel dont l'allure de l'impédance d'entrée est proche de $Z_l$ \cite{}. 

\par En considérant des cheminées relativement courtes (ce qui exclut le trou de registre par exemple), les paramètres $s$ et $m$ caractérisent les cellules :
\end{comment}
\begin{equation}
\label{eq:param}
%m=\frac{r_t}{r}\frac{1}{\sqrt{2sh_c}} \Rightarrow
Z_l = jZ_c\tan{(ks)}\sqrt{\frac{m^2s^2+ks\cot{(ks)}}{m^2s^2-ks\tan{(ks)}}}
\mbox{ où }
m=\frac{r_t}{r}\frac{1}{\sqrt{2sh_c}}
\end{equation} 
Dans l'expression (\ref{eq:param}) apparaissent une infinité de coupures qui définissent des bandes fréquentielles aux comportements variés :
\begin{itemize}
\item en basses fréquences, jusqu'à la coupure proche de $f_c\simeq mc/(2\pi)$ (la plus faible pulsation solution de $m^2s^2=ks\tan{ks}$) de l'ordre de $1$ à $2\kilo\hertz$, le réseau se comporte comme un élément de longueur équivalente $s'=s\sqrt{1+(ms)^{-2}}$. L'hypothèse de troncature du résonateur au niveau du premier trou ouvert reste valable dans cette bande de fréquence et pour des cheminées courtes et/ou larges ($ms$ grand). Dans le cas contraire, les ondes stationnaires basses fréquences s'étendent sur une partie du guide en aval du premier trou.
\item au delà, jusqu'à $k={\pi}/{(2s)}$ soit jusqu'aux environs de $4$ à $5\kilo\hertz$, se trouve la première bande passante du réseau.
\item on observe ensuite, pour le modèle décrit par l'équation~(\ref{eq:param}), un succession de bandes d'arrêt et de bandes passantes\ldots
\end{itemize}
\par Les considérations précédentes reposent sur l'hypothèse que les cellules sont identiques. En pratique, on peut montrer que si chaque cellule a la même fréquence de coupure, à condition que ses dimensions soient petites devant la longueur d'onde, le réseau se comporte exactement comme un réseau périodique. Benade \cite{Ben76} a constaté que la fréquence de coupure change très peu d'un doigté à l'autre (hors doigtés de fourche), ce qui entraîne l'homogénéité de timbre. Ceci semble témoigner d'un réseau parfaitement périodique. Pourtant, afin d'assurer une progression logarithmique en fréquence des notes, il faut écarter les trous de plus en plus quand on s'éloigne de l'anche, et on constate que le diamètre des trous croît également, assurant une fréquence de coupure à peu près constante.

\begin{comment}
le plus important est que chaque élément introduise la même première fréquence de coupure. Si les trous s'éloignent de plus en plus pour assurer la justesse de l'instrument, on constate sur la clarinette réelle qu'ils deviennent également plus larges : la fréquence de coupure est sensiblement constante. C'est la similitude d'un point de vue acoustique qui est importante, plus l'exacte périodicité géométrique. L'homogénéité du son est assurée même si les cellules ne sont pas rigoureusement identiques.
\end{comment}

\subsection*{Le résonateur complet}
\label{ssec:resonateur}
Les pics de résonance du résonateur sont modifiés par la présence du réseau de trous latéraux ouverts placé à l'extrémité d'un guide cylindrique :
\begin{itemize}
\item ceux qui se trouvent dans les bandes d'arrêt du réseau sont légèrement déplacés vers les basses fréquences puisque les ondes stationnaires s'établissent sur la longueur du cylindre augmentée d'une correction fonction des paramètres géométriques des ouvertures et de l'écartement entre les trous,
\item ceux qui correspondent à la bande passante du réseau sont fortement atténués car une grande partie de l'énergie est prélevée par le réseau : les résonances ne peuvent plus s'établir.
\end{itemize}
%\begin{figure}[hbt]
%\begin{center}
%\includegraphics[width=9cm]{EffetReseauAbs3.eps}
%\includegraphics[width=9cm]{EffetReseauAngle3.eps}
%\end{center}
%\caption{Effet du réseau de trous ouverts : impédance d'entrée d'un cylindre ($L=0.5\metre$, $r=7\milli\metre$) suivi d'un réseau de trous ouverts ($s=1\centi\metre$, $m=43\reciprocal\metre$) en pointillés et simple cylindre (de longueur corrigée $L+s'$) en trait plein.}
%\label{fig:effetreseau}
%\end{figure}

\section*{Mise en \oe uvre numérique}
\label{sec:numerique}

%\subsection*{Effet passe-bas}
\par En se focalisant sur la première bande d'arrêt et sur la première bande passante, il est possible d'assimiler --~dans un domaine de fréquence de l'ordre de $0-5000\hertz$~-- le réseau vu par le tuyau cylindrique à un système absorbant les hautes fréquences, c'est-à-dire un filtre passe-bas pour le coefficient de réflexion présenté par le réseau au cylindre. C'est cet effet que nous avons cherché à retranscrire de manière adaptée aux méthodes numériques de synthèse sonore. Une exigence qui accompagne le processus de synthèse sonore est le choix de la simplicité des modèles adoptés. Nous nous imposons l'emploi de filtres numériques d'ordres faibles mais dont les coefficients sont reliés aux paramètres géométriques et acoustiques du réseau.
\par Il est important de noter que la principale difficulté provient du fait que, à cause de la forme de l'équation~\ref{eq:param}, les résonances du réseau considéré ne sont pas des résonances classiques auxquelles sont associées un nombre fini de pôles dans le plan complexe. 

\par
\begin{figure}[hbt]
\begin{center}
\includegraphics[width=9cm]{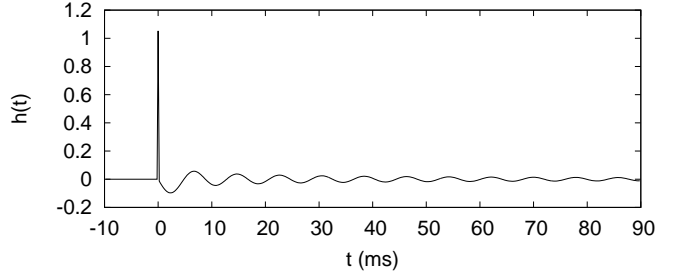}
\end{center}
\caption{Réponse impulsionnelle $h(t)$ associée à l'impédance d'entrée $Z_l(\omega)$ (équation (\ref{eq:param}) ) du réseau de trous latéraux.}
\label{fig:dirac}
\end{figure}

 Le tracé de la réponse impulsionnelle $h(t)$ associée à l'impédance réduite du réseau seul indique (Figure~\ref{fig:dirac}) la présence d'une discontinuité à l'instant initial $t=0$ et d'oscillations décroissantes, ce qui rend pertinent la décomposition de $h(t)$ en la somme d'une impulsion de Dirac de hauteur $h_0$ et d'une fonction sinusoïdale amortie :
\begin{gather}
\label{eq:decompo}
\tilde{h}(t)=h_0\delta(t)-Ae^{-\beta t}\sin{(\omega_0 t)}\\
\tilde{Z_l}(j\omega)=h_0-%
\frac{A}{1+q_r\frac{j\omega}{\omega_r}+\left(\frac{j\omega}{\omega_r}\right)^2}
\end{gather}
\vspace{-5mm}
\par La contribution à l'instant initial de la partie oscillante semblant nulle, la hauteur $h_0$ peut être évaluée analytiquement en fonction des paramètres $m$ et $s$. Les coefficients restants sont déterminés pour que le filtre présente le même comportement dans les deux premières bandes que l'impédance itérative du réseau. De plus, le réseau étant un élément passif (il n'est pas source d'énergie), le coefficient de réflexion doit présenter un module inférieur à l'unité. On peut montrer que ceci impose $q_r\geq 1$. Si l'on note $\lambda$ les pertes visco-thermiques à la pulsation de coupure $mc$, les conditions précédentes mènent aux valeurs suivantes :
\begin{eqnarray}
h_0 &=& A \ \ =\ \ f(m,s)\,,\\
q_r &=& 1\,,\\
\omega_r &=& c\sqrt{m^2+\lambda^2}\,.
\end{eqnarray}
Ce modèle analogique est ensuite discrétisé à la fréquence d'échantillonnage $F_e=44100\hertz$ à l'aide de schémas numériques centrés :
\begin{eqnarray}
j\omega &\longleftrightarrow &Dz=\frac{F_e}{2}(z-z^{-1})\,\\
(j\omega)^2 &\longleftrightarrow &D^2z={F_e}^{2}(z-2+z^{-1})\,.
\end{eqnarray}
\begin{figure}[hbt]
\begin{center}
\includegraphics[width=9cm]{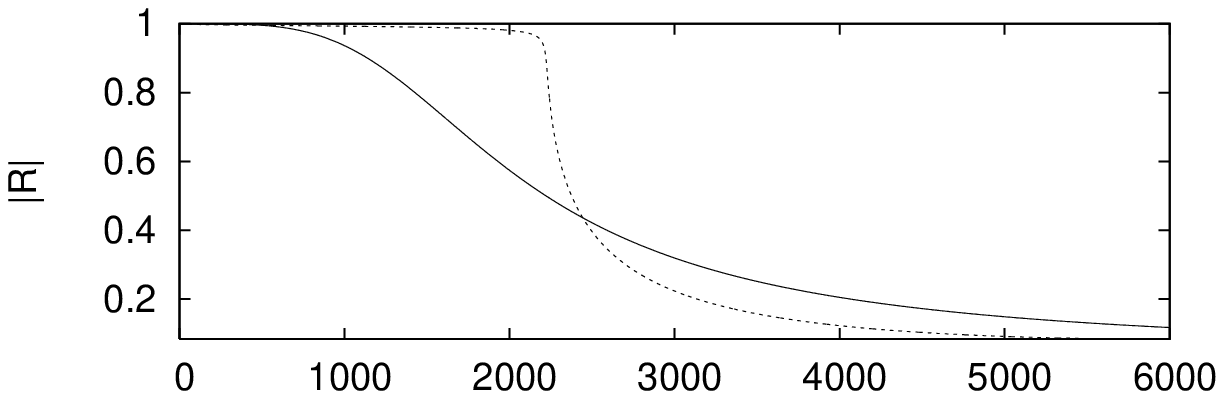}
\includegraphics[width=9cm]{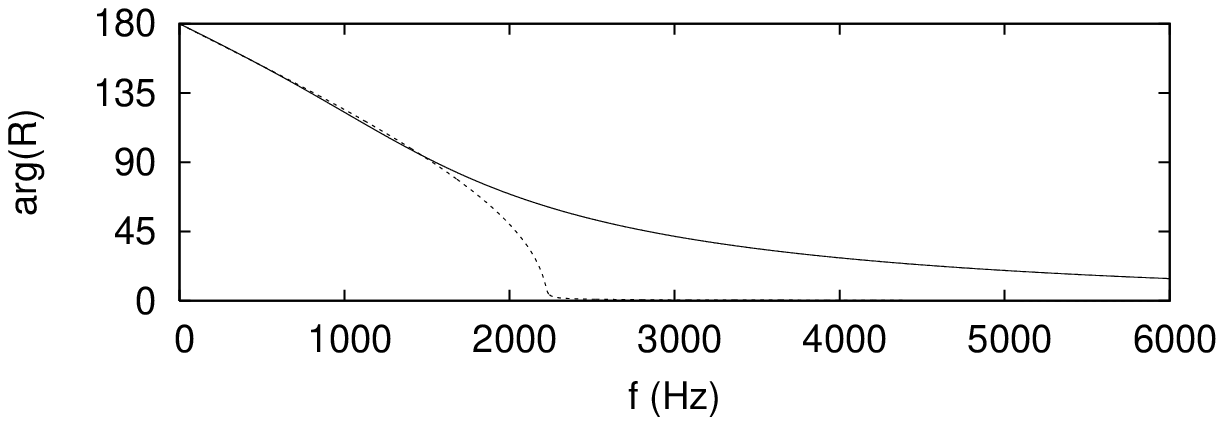}
\end{center}
\caption{Coefficient de réflexion en entrée du réseau : modèle de réseau (en pointillés) et filtre numérique approché (en trait plein). $r=7\milli\metre$, $s=1\centi\metre$ et $m=43\reciprocal\metre$.}
\label{fig_coefrefl}
\end{figure}
\begin{figure}[hbt]
\begin{center}
\includegraphics[width=9cm]{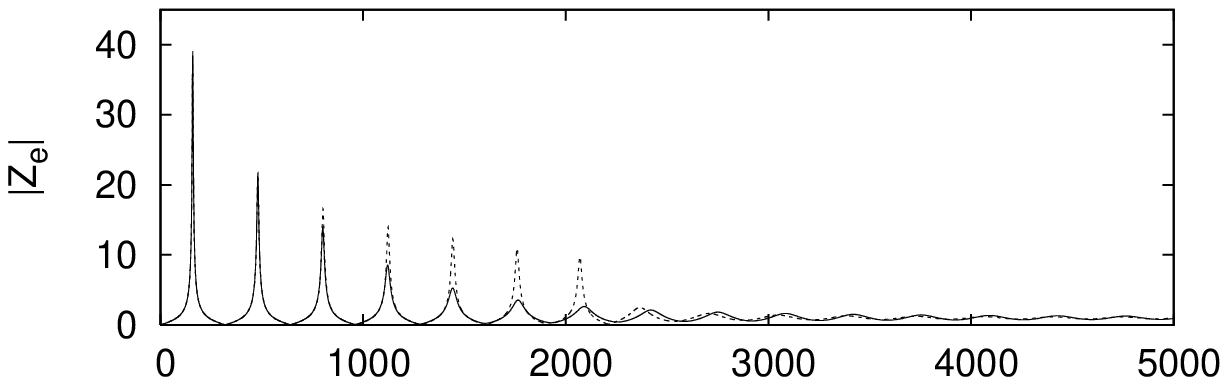}
\includegraphics[width=9cm]{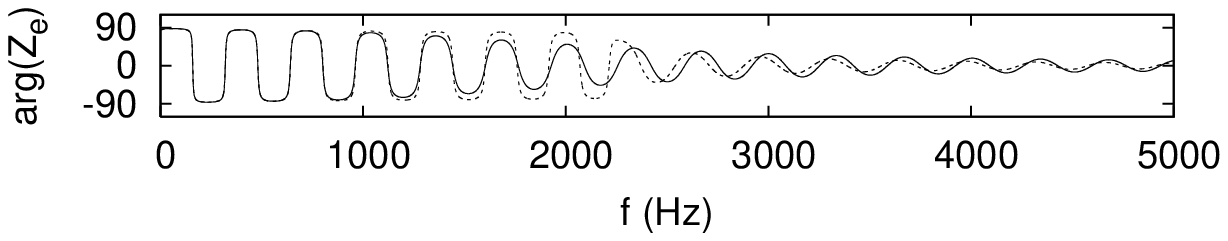}
\end{center}
\caption{Impédances d'entrée du résonateur complet avec l'impédance itérative $Z_l(\omega)$ (en pointillés) et avec la mise en \oe uvre numérique $\hat{Z_l}(z)$ (en trait plein). $L=0.5\metre$, $r=7\milli\metre$, $s=1\centi\metre$ et $m=43\reciprocal\metre$.}
\label{fig:effetreseau}
\end{figure}
On obtient ainsi un filtre numérique $\hat{Z_l}(z)$ introduisant une atténuation des pics de résonance du tube cylindrique au-dessus de la fréquence de coupure. La présence d'un pic d'impédance du réseau correspondant à une coupure très raide pour le coefficient de réflexion (Figure~\ref{fig_coefrefl}) nécessiterait, pour être prise en compte plus précisément, l'emploi d'un filtre numérique d'ordre très élevé. Le filtre retenu restitue cependant relativement bien le comportement du réseau dans les premières bande d'arrêt et bande passante.

\par On prend ensuite en compte le comportement acoustique du tuyau qui prend être modélisé comme un retard associé à un filtre numérique du 1° ordre modélisant la propagation avec pertes visco-thermiques \cite{GKV05}. Il est alors possible de décrire numériquement le résonateur complet par une équation aux différences -- dont les coefficients sont reliés aux différents paramètres géométriques -- de la forme :
\begin{equation}
p_e(t_n) = u_e(t_n) +V
\end{equation}
où $V$ est fonction d'un petit nombre de valeurs passées et connues de la pression et du débit d'entrée.
\begin{comment}
\par La présence d'une résonance d'un type inhabituel nous a obligé à faire un choix de modélisation pour la représentation numérique obéissant à une volonté de simplicité liée à l'aspect temps-réel de la synthèse.  Cependant, au voisinage de la fréquence de coupure, les exigences de stabilité (non-amplification des ondes par le réseau) imposent une résonance bien moins marquée que celle décrite dans la première partie. Il serait possible d'augmenter l'ordre du filtre numérique. Néanmoins, il semble que la hauteur du pic de résonance soit moins importante que les effets de correction de longueur (dans la bande d'arrêt) et d'absorption des ondes entrantes (dans la bande passante) lorsque l'on s'intéresse à l'impédance d'entrée du résonateur complet, ce qui justifie le choix de se limiter à un filtre d'ordre 2.
\end{comment}

\section*{Optimisation}
\label{sec:optimisation}
%Une fois le modèle élaboré, il reste à déterminer quelles valeurs donner aux divers paramètres. 
Les valeurs des différents paramètres du modèle complet sont déterminés par comparaison à une référence. 
Nous disposons d'une série de mesures d'impédances d'entrée de clarinette pour une série de doigtés du registre chalumeau, réalisées en chambre sourde au Laboratoire d'Acoustique de l'Université du Maine. On dispose ainsi des valeurs associées au mode plan \cite{DB92} pour un ensemble de 300 fréquences entre $100$ et $2500\hertz$. L'objectif est d'ajuster la longueur $L$ et le rayon $r$ du tuyau ainsi que les paramètres $m$ et $s$ du réseau de sorte que le filtre numérique $\hat{Z_l}(z)$ élaboré à partir de ces valeurs coïncide avec la référence expérimentale $Z_{ref}$ pour les pulsations $\omega_n$ considérées. Pour cela, nous avons eu recours à des méthodes d'optimisation globale comme les méthodes de recuit simulé avec une fonction coût de la forme
%\cite{Ing89}
\begin{equation}
\label{eq:fctcout}
F(l,r,m,s) = \sum_{n=1}^N f(\omega_n)
	\left|\hat{Z_l}\left(e^{\displaystyle j\frac{\omega_n}{F_e}}\right)%
	-Z_{ref}(\omega_n)\right|^2.
\end{equation}
$f(\omega)$ est une fonction de pondération qui permet d'accorder une pénalité accrue aux écarts entre les pics d'impédance obtenus expérimentalement et ceux avec le modèle numérique, sanctionnant ainsi un mauvais ajustement en fréquence et/ou en amortissement des résonances du système.
\par L'optimisation aboutit à des valeurs de $L$ et $r$ relativement proches de celles mesurées sur la clarinette utilisée pour les mesures d'impédance, mais aussi à des résultats pour les paramètres du réseau de trous ouverts qui sont cohérents avec les dimensions et l'écartement des ouvertures du corps de la clarinette. Ceci semble confirmer que le filtre numérique $\hat{Z_l}$ retranscrit de manière efficace l'effet du réseau de trous ouverts.

\section*{Conclusion}
\label{sec:conclusion}
\par Nous avons, à l'heure actuelle, déjà pu comparer les auto-oscillations obtenues à partir du modèle cylindrique, du modèle numérique complet (cylindre et réseau) et des impédances mesurées (en utilisant la technique décrite dans \cite{GGA95}). D'un point de vue du timbre, les sons correspondant au modèle incluant le réseau sont beaucoup plus proches des sons synthétisés avec les impédances mesurées que ceux obtenus à partir d'un simple cylindre. En revanche, nous n'avons pu comparer le fonctionnement en auto-oscillation pour le filtre numérique adopté pour le réseau et pour un filtre présentant une coupure bien plus marquée. Il faudra de plus s'intéresser de manière théorique à l'influence de l'atténuation des pics situés au-dessus de la fréquence de coupure du réseau sur le fonctionnement global de l'instrument, notamment sur les régimes non-stationnaires. D'un point de vue de la synthèse, l'étape suivante pourra consister en l'élaboration d'un modèle de rayonnement corrélé au modèle de réseau de trous latéraux ouverts.
\begin{comment}
Il faudra s'intéresser à l'influence de l'atténuation des pics situés au-dessus de la coupure d'un point de vue du fonctionnement global de l'instrument. En particulier, comment les régimes non-stationnaires et les attaques en particulier dépendent de l'amplitude des pics supérieurs du résonateur.
\end{comment}
%\section*{Remerciements}
\par Ce travail a été effectué dans le cadre de \textsc{Consonnes}, projet soutenu par l'Agence Nationale de la Recherche.
%\vspace{-2mm}

\end{document}